\documentclass[runningheads]{llncs}
\usepackage[T1]{fontenc}

\usepackage{amsmath}
\usepackage{amsfonts}
\usepackage{bbm}

\usepackage{natbib}
\usepackage{graphicx}
\usepackage{float}
\usepackage{subcaption}
\usepackage{threeparttable}
\usepackage{hyperref}
\usepackage{enumitem}
\usepackage{multirow}
\usepackage{booktabs}
\usepackage{rotating}
\usepackage{placeins}
\usepackage{caption}
\usepackage{colortbl}
\usepackage{makecell}

\DeclareMathOperator*{\argmax}{arg\,max}

\usepackage[dvipsnames]{xcolor}

\begin{document}
\title{Embedding Cultural Diversity in Prototype-based Recommender Systems}
%
% \titlerunning{Embedding Cultural Diversity in }
\author{Armin Moradi \inst{1, 2} \and
Nicola Neophytou \inst{1} \and
Florian Carichon \inst{1,3} \and
Golnoosh Farnadi\inst{1, 3}}
\authorrunning{Moradi et al.}
\institute{Mila - Quebec AI Institute \and
Université de Montréal \and
McGill University
}
\maketitle
\begin{abstract}

Popularity bias in recommender systems can increase cultural overrepresentation by favoring norms from dominant cultures and marginalizing underrepresented groups. This issue is critical for platforms offering cultural products, as they influence consumption patterns and human perceptions. In this work, we address popularity bias by identifying demographic biases within prototype-based matrix factorization methods. Using the country of origin as a proxy for cultural identity, we link this demographic attribute to popularity bias by refining the embedding space learning process. First, we propose filtering out irrelevant prototypes to improve representativity. Second, we introduce a regularization technique to enforce a uniform distribution of prototypes within the embedding space. Across four datasets, our results demonstrate a 27\% reduction in the average rank of long-tail items and a 2\% reduction in the average rank of items from underrepresented countries. Additionally, our model achieves a 2\% improvement in HitRatio@10 compared to the state-of-the-art, highlighting that fairness is enhanced without compromising recommendation quality. Moreover, the distribution of prototypes leads to more inclusive explanations by better aligning items with diverse prototypes.

% Cultural overrepresentation in recommender systems can perpetuate inequality by favoring norms from dominant cultures and marginalizing underrepresented groups. This issue is critical for platforms offering cultural products, as they influence consumption patterns and human perceptions. In this work, we identify demographic biases present in prototype-based matrix factorization methods used in recommender systems with cultural applications. By linking demographic and popularity biases to properties of the embedding space, we propose two enhancements: (i) filtering out irrelevant prototypes to improve generalizability, and (ii) introducing a regularization technique to enforce a uniform distribution of prototypes within the embedding space. On the LastFM dataset, our model achieves a HitRatio@10 of 0.600—a 3\% improvement over the state-of-the-art—and reduces the average rank of long-tail items by 38\%. We also observe a 3\% lower average rank for items from underrepresented countries without compromising recommendation quality. Furthermore, distributing prototypes generates more inclusive explanations by aligning items with diverse prototypes.\footnote{Our code will be made publicly available upon publication of this work.}

\end{abstract}

\section{Introduction}
\label{sec:introduction}

% Paragraph 1 (Motivation & Socio-Technical Issue): What is culture? (or what we mean by culture?), and how is it related to recommendation systems (Recsys)? What impact do Recsys have on culture, and how does this connect to fairness? How does such bias lead to disparities and result in the exclusion of minority and marginalized communities’ cultures from representation?

Culture represents the shared beliefs, values, norms, and practices of particular societies or groups \citep{spencer2012culture}. In today's digital world, recommender systems of platforms such as Netflix, Amazon, Steam, and Spotify have become the primary, and sometimes the only, means of accessing, discovering, and consuming cultural content and products \citep{ferraro2022measuring}. These systems play a significant role in shaping cultural consumption patterns, influencing the selective retention and transmission of cultural artifacts \citep{brinkmann2023machine}. However, a major issue arises when these systems are biased—particularly toward popular products—leading to \emph{popularity bias} \cite{wei2021model}. This bias can amplify dominant cultural norms, marginalizing underrepresented groups and their cultural expressions \citep{shuker2012popular, lesota2022traces, bello2021cultural}, resulting in the exclusion of minority communities from fair representation \citep{deldjoo2024fairness}. The reinforcement of dominant values can perpetuate stereotypes \cite{song2013deconstruction, guerlain1997ironies}, resulting in representational harm, while allocational harm arises as opportunities for minority communities' cultural products remain limited, thus exacerbating socio-economic disparities \citep{spring2016deculturalization}.

% Paragraph 2 (Technical Gap): How do cultural issues in Recsys relate to the underlying methodology, such as matrix factorization (MF)? Why do these issues arise, and how have they been addressed in the literature? (you can connect it to fairness in Recysts, long-tail and popularity bias here)

Biases in recommendation algorithms are rooted in the design and limitations of widely adopted techniques, such as matrix factorization (MF) and deep learning-based models \citep{zhang2019deep}. These approaches excel at capturing complex user-item interactions but often fail to address fairness, as their underlying latent factor models emphasize patterns driven by popularity \citep{koren2009matrix}. Prototype-based recommender systems, such as Prototype-based Matrix Factorization (ProtoMF) \citep{melchiorre2022protomf}, were introduced to enhance explainability by identifying prototypes representing typical behaviors. However, these systems have not fully addressed fairness concerns—particularly how prototypes might still reflect dominant cultural norms, thus perpetuating biases in cultural representation.

% Paragraph 3 (Our Technical Solution): How does our approach differ from previous work? In what ways does considering the embedding space help to resolve the gaps and issues identified in Paragraph 1? Also, how can our approach be used in the prototype MF by providing better explanations for various cultures? (you need to provide low-level technical novelty here rather than high-level claims)

To tackle these limitations, we introduce a novel approach that improves ProtoMF models with two key innovations: 
\vspace{-0.5cm}
\begin{itemize}
    \item \emph{Prototype K-filtering}, which refines the representation of users and items by selecting the $k$ nearest prototypes, allowing the model to better capture underrepresented items and users; 
    \item  \emph{Prototype-Distributing Regularizer}, a mechanism that encourages a more even distribution of prototypes across the embedding space, ensuring diverse cultural representations by reducing the clustering effect around popular items.
\end{itemize}
\vspace{-0.5cm}
We address demographic bias by incorporating the country of origin of item providers as a cultural feature, thereby tackling popularity bias through a demographic lens. This approach enables us to mitigate bias while maintaining model performance.

% Paragraph 4: A list of our contributions, with references to the relevant sections. You need to include both the extensions to the prototype MF and the evaluation results across various datasets. You can include some interesting results to motivate the reader to read more. (edited) 

We evaluate our framework on MovieLens-1M \citep{harper2015movielens}, LastFM-2b \citep{MELCHIORRE2021102666}, and Amazon Reviews'23 \citep{hou2024bridging} for `Musical Instruments' and `Beauty and Personal Care' categories; due to their strong ties to cultural behavior. Our extensive experiments assess both the performance and fairness of the recommendations. The results indicate that demographic bias in prototype-based recommender systems can be effectively mitigated without compromising utility. 

The remainder of this paper is organized as follows: Section \ref{sec:related_work} reviews related work, Section \ref{sec:preliminaries} presents background on ProtoMF, and Section \ref{sec:methodology} details our proposed methodology. Section \ref{sec:evaluation_setup} describes the evaluation setup, and Section \ref{sec:results} discusses the results. Finally, we provide a detailed discussion in Section \ref{sec:discussion}.

\section{Related Work}
\label{sec:related_work}

% We begin by reviewing existing literature to situate our work within the broader research landscape of cultural recommender systems (RecSys), the role of fairness in these systems, existing solutions, and how prototype-based methods can offer an explainable fairness strategy.

Culture plays an important role in shaping user preferences and interactions in recommender systems (RecSys). Several studies have explored how the cultural distance between users and items influences satisfaction, and they emphasize the need to integrate cultural factors into RecSys design \citep{moreau2004cultural}. Existing work has incorporated users' cultural backgrounds into collaborative filtering models by embedding demographic features into profiles and interactions \citep{zangerle2018culture}. \citet{hong2021multi} expanded this by embedding cultural factors directly into user profiles, while \citet{hong2024cross} grouped users based on shared cultural backgrounds to improve recommendation accuracy. However, these works mainly focused on improving recommendation utility rather than addressing biases, particularly when algorithms favor dominant cultures \citep{brinkmann2023machine}. \citet{ferraro2022measuring} highlighted this issue by introducing a measure of cultural diversity in collaborative filtering systems, aligning with \citet{lesota2022traces}'s findings related to music RecSys. Building on these insights, our work explores mitigating popularity bias in RecSys to guarantee a better representation of underrepresented cultural groups, shifting the focus from utility alone to fairness and diversity in recommendations.

Fairness in RecSys presents a multifaceted challenge, especially when popularity and demographic biases are involved \citep{sonboli2022multisided}. These biases can result in representational harms, such as reinforcing stereotypes, and allocational harms, where opportunities for certain groups or products are limited \citep{barocas2023fairness}. Popularity bias, in particular, often skews recommendations toward dominant subgrouups, leading to the underrepresentation and marginalization of less popular ones \cite{lesota2021analyzing}. This can be investigated from both the users and the item providers' perspectives (\cite{sonboli2021fairness}). Our work specifically focuses on the item dimension for two reasons: first, by targeting items’ country of origin, we can directly mitigate allocational harms affecting underrepresented cultures; second, current datasets often contain sensitive information on items rather than users, allowing us to identify and challenge strong biases. By mitigating popularity bias through this lens, we aim to improve the representation of underrepresented cultural products, promoting fairness across a broader cultural spectrum.

Several fairness strategies have been proposed to address popularity and demographic biases, generally categorized into pre-processing, in-processing, and post-processing methods \citep{deldjoo2024fairness}. For instance, \citet{rhee2022countering} use an in-processing method that mitigates popularity bias by adding a regularizer to minimize score differences, while \citet{wei2021model} employ a counterfactual method to reduce popularity bias. \citet{zheng2021disentangling} address the same issue by assigning separate embeddings for user interest and conformity. Also, \cite{beutel2019fairness} add a regularizer to prevent favoring of one group over another by penalizing prediction discrepancies between groups. In our work, we also propose an in-processing mitigation method. However, we distinguish ourselves by modifying the learning process that changes the structure of the embedding space without conflicting with the model’s downstream task.

Our focus on the embedding space is enabled by prototype-based methods, which use prototypes as anchor points to generate explainable embeddings for both items and users \citep{sankar2021protocf}. By introducing bias-mitigation techniques within \citeauthor{melchiorre2022protomf}'s prototype-based matrix factorization (ProtoMF), we position ourselves among the few proposing explainable fairness strategies through the learning process of the embedding space, which is shown to align with improving providers' fairness \cite{dinnissen2023control}. Unlike \citet{melchiorre2024modular}, who focus on unbiasing ProtoMF using user-related coefficients, our approach modifies the learning process to create a more inclusive embedding space, aiming to balance the representations and provide fairer outcomes for item providers.

\section{Prototype-based Matrix Factorization} \label{sec:preliminaries}

Prototype-based Matrix Factorization (ProtoMF) extends traditional matrix factorization by introducing prototypes to redefine the embedding process for both users and items. Specifically, we define two sets of prototype vectors:

\begin{itemize}
    \item User prototypes: $P^u = {p^{u}_{1}, p^{u}_{2}, \ldots, p^{u}_{L_u}} \subset \mathbb{R}^{d}$, forming a matrix $P^u \in \mathbb{R}^{L_u \times d}$.
    \item Item prototypes: $P^i = {p^{i}_{1}, p^{i}_{2}, \ldots, p^{i}_{L_i}} \subset \mathbb{R}^{d}$, forming a matrix $P^i \in \mathbb{R}^{L_i \times d}$.
\end{itemize}

Here, $L_u$ and $L_i$ are the numbers of prototypes for users and items, respectively, and $d$ is the dimensionality of the latent space. These prototype vectors serve as anchor points to create more interpretable embeddings \citep{melchiorre2022protomf}. The embeddings of users and items are transformed using these prototypes. For user and item embeddings $u, i \in \mathbb{R}^d$, the transformed embedding $u^* \in \mathbb{R}^{L_u}, i^* \in \mathbb{R}^{L_i}$ is computed as:
\begin{equation} \begin{split}
     u^* = \left[ \text{sim}(u, p^{u}_{1}), \ldots, \text{sim}(u, p^{u}_{L_u}) \right], \
      i^* = \left[ \text{sim}(i, p^{i}_{1}), \ldots, \text{sim}(i, p^{i}_{L_i}) \right]. 
\end{split}\end{equation}

\noindent where $\text{sim}(\cdot, \cdot)$ denotes the shifted cosine similarity.

This process yields transformed user and item matrices $U^* \in \mathbb{R}^{N \times L_u}$ and $I^* \in \mathbb{R}^{M \times L_i}$, where $N$ and $M$ are the numbers of users and items, respectively.

The affinity score between a user and an item is calculated as:

\begin{equation}
    \text{Aff}(u, i) = {{u}^*}^\top {\hat{i}} +{\hat{u}}^\top{{i}^*},
    \label{eq:transformed_vectors}
\end{equation}

\noindent where $\hat{u} = W^u u \in \mathbb{R}^{L_i}$ and $\hat{i} = W^i i \in \mathbb{R}^{L_u}$ are linear transformations mapping the embeddings to the opposite prototype spaces.

The recommendation loss for users ($L_{U-rec}$) and items ($L_{I-rec}$) is defined using implicit feedback as in \cite{rendle2021item}, by applying the softmax function over the affinity scores across all possible items (for users) and all possible users (for items).

In addition to the recommendation loss, two collaborative regularization terms are introduced for both the user and item sides. On the user side, the first regularization term (Eq. \ref{eq:original_regularizer1}) increases the similarity between each prototype $p_l^u$ and its most similar user $u$, and the second term (Eq. \ref{eq:original_regularizer2}) increases the similarity between each user $u$ and its most similar prototype:

\begin{equation}
      R_{\{P^u \rightarrow u\}} = - \frac{1}{L^u}\sum_{l=1}^{L^u} \max_{l \in [1,\ldots, N]} \text{sim}(u, p_l^u),
      \label{eq:original_regularizer1}
\end{equation}

\begin{equation}
    R_{\{u \rightarrow P^u\}} = - \frac{1}{N}\sum_{l=1}^{N} \max_{l \in [1,\ldots, L^y]} \text{sim}(u_i, p_l^u).
    \label{eq:original_regularizer2}
\end{equation}

The final user and item loss functions are calculated by the summation of the recommendation loss, and the two regularization factors. Formally, the loss functions for the user side is: $L_{U-proto} = L_{U-rec} + \lambda^{u}_{1}R_{\{P^u \rightarrow u\}} + \lambda^{u}_{2}R_{\{u \rightarrow P^u\}}$, where the set of $\lambda_1^u$ and $\lambda_2^u$s control the weights of the regularization terms.

The item side regularization terms and loss function are defined analogously. Therefore, the total loss function is the sum of the user and item loss functions: 
\begin{equation}
L_{\text{total}} = L_{U-proto} + L_{I-proto}.    
\end{equation}

\section{Methodology} \label{sec:methodology}

The following subsections detail the two enhancements and their integration into the ProtoMF framework.

\subsection{Prototype K-filtering}

In the standard ProtoMF model, all prototypes contribute to computing the transformed user and item representations $u^*$ and $i^*$. This means that even distant prototypes can influence these representations, potentially biasing the model toward popular items and users. To address this issue, we introduce \textit{Prototype K-filtering}, which retains only the $k$ nearest prototypes for each user and item, setting the similarities to the remaining prototypes to zero.

Formally, for a user $u$ and item $i$, we select the $k_u \ll L_u$ and $k_i \ll L_i$ indices of the nearest prototypes:

\begin{equation}
 P^{u}_k = \argmax_{P^{u} \subset [L_{u}]: |P^{u}| = k_u } \sum_{p^{u}_{j} \in P_u} sim(u,p^{u}_{j}),
\end{equation}

\begin{equation}
 P^{i}_k = \argmax_{P^{i} \subset [L_{i}]: |P^{i}| = k_i } \sum_{p^{i}_{j} \in P_i} sim(i,p^{i}_{j}).
\end{equation}

The filtered transformed representations are computed as: $u^{*, k} = u^* \odot \mathbbm{1}_{P^{u}_k}$ for users and $\ i^{*, k} = i^* \odot \mathbbm{1}_{P^{i}_k}$ for items, where $\odot$ denotes element-wise multiplication, and $\mathbbm{1}_{P^{u}k}$ and $\mathbbm{1}_{P^{i}_k}$ are indicator vectors with ones at positions corresponding to the selected prototypes and zeros elsewhere.

By focusing on local information, this method improves the representation of less popular users and items, reducing the influence of dominant prototypes that may contribute to bias. In our experimental results, models using k-filtering are denoted as $\{\text{User}\}_{\{k\}}$ when k-filtering is applied only to the user side, $\{\text{Item}\}_{\{k\}}$ when applied only to the item side.

\subsection{Prototype-Distributing Regularizer}

Another concern is that prototypes may cluster in specific regions of the embedding space, causing underrepresented users or items to be associated with distant prototypes. To mitigate this, we introduce a regularization term that encourages prototypes to be uniformly distributed across the embedding space.

The modified loss function becomes:

\begin{equation}
       L = L_{\text{total}} + \lambda_u \cdot ||\hat{P}_{L_u} \hat{P}_{L_u}^T||_F + \lambda_i \cdot ||\hat{P}_{L_i}  \hat{P}_{L_i}^T||_F,
\end{equation}

\noindent where $L_{\text{total}}$ is the original ProtoMF loss,
$\hat{P}^{u} \in \mathbb{R}^{L_u \times d}$ and $\hat{P}^{i} \in \mathbb{R}^{L_i \times d}$ are matrices of normalized user and item prototypes (each row vector normalized to unit length),
$I$ is the identity matrix,
$||\cdot||_F$ denotes the Frobenius norm,
$\lambda_u$ and $\lambda_i$ are hyperparameters controlling the strength of the regularization.

This regularizer encourages the prototypes to be approximately orthogonal, promoting a uniform distribution that better covers the embedding space. By preventing prototypes from clustering, we ensure a more diverse representation of user preferences and item characteristics, reducing bias toward dominant groups. In our experimental results, models using the distributing regularizer are denoted as $\{\text{User}\}_{\{\lambda\}}$ when the regularizer is applied only to the user side, $\{\text{Item}\}_{\{\lambda\}}$ when applied only to the item side.

When both k-filtering and the distributing regularizer are applied to both sides, we denote the model as $\{\text{User-Item}\}_{\{k, \lambda\}}$.

\section{Evaluation Setup} \label{sec:evaluation_setup}

\subsection{Datasets}

We conduct our experiments using four datasets:
\begin{itemize} 
\item LastFM-2b \citep{MELCHIORRE2021102666}: A subset augmented with artists' countries of origin by matching artist names with the \href{musicbrainz.org}{MusicBrainz} database. 
\item MovieLens-1M \citep{harper2015movielens}: Users' movie ratings, enriched with movies' countries of origin retrieved via \href{https://www.themoviedb.org/}{The Movie Database} API. 
\item Amazon Reviews'23 \citep{hou2024bridging}: We use two categories—Musical Instruments, Beauty and Personal Care, filtered to include items' countries of origin. These categories are selected because cultural factors, such as differing beauty standards between Asian and European countries \citep{frederick2015beauty}, can influence user decisions.
\end{itemize}

To simplify the evaluation, we define groups based on interaction counts and the country of origin as the sensitive attribute: 1. \textit{Overrepresented groups}: Countries in the top 10\% by number of interactions. 2. \textit{Underrepresented groups}: Countries in the third quartile (25\%-50\%) of interaction counts. We exclude the extreme tail of the distribution to balance representation and statistical reliability. Dataset statistics and groupings are detailed in Table~\ref{tab:datasets_combined}.

\vspace{-0.7cm}
\begin{table}[h]
\centering
\caption{Overview of datasets before and after filtering and the choice of underrepresented and overrepresented groups.}
\label{tab:datasets_combined}
\begin{subtable}[t]{\textwidth}
    \centering
    \caption{Overview of datasets before and after filtering based on the sensitive attribute of country of origin.}
    \label{tab:datasets_1}
    \resizebox{\textwidth}{!}{
    \begin{tabular}{|c|c|c|c|c|c|c|}
    \hline
    \multirow{2}{*}{\textbf{Dataset}} &
      \multicolumn{3}{c|}{\textbf{Before Filtering}} &
      \multicolumn{3}{c|}{\textbf{After Filtering}} \\ \cline{2-4} \cline{5-7} 
     &
      \multicolumn{1}{c|}{\textbf{\#users}} &
      \multicolumn{1}{c|}{\textbf{\#items}} &
      \textbf{\#interactions} &
      \multicolumn{1}{c|}{\textbf{\#users}} &
      \multicolumn{1}{c|}{\textbf{\#items}} &
      \textbf{\#interactions} \\ \hline
    Beauty & 11.3M & 1.0M & 23.9M & 508K ($\downarrow 95.5\%$) & 35.8K ($\downarrow 96.4\%$) & 1.5M ($\downarrow 93.7\%$) \\ \hline
    % Grocery& 7.0M & 603.2K & 14.1M & 340K ($\downarrow 95.1\%$) & 23.3K ($\downarrow 96.1\%$) & 1.0M ($\downarrow 92.9\%$) \\ \hline
    Instruments & 1.8M & 213.6K & 3.0M & 9.9K ($\downarrow 95.8\%$) & 9.4K ($\downarrow 95.5\%$) & 212.3K ($\downarrow 92.9\%$) \\ \hline
    % Video Games & 2.8M & 137.2K & 4.6M & 32.4K ($\downarrow 98.8\%$) & 4.1K ($\downarrow 97.0\%$) & 79.6K ($\downarrow 98.2\%$) \\ \hline
    LastFM-2b & 15K & 4.0M & 30.3M & 8.4K ($\downarrow 44\%$) & 63.0K ($\downarrow 98.4\%$) & 1.48M ($\downarrow 95.1\%$) \\ \hline
    MovieLens-1M & 6K & 4K & 1M & 6K ($\downarrow 0\%$) & 4K ($\downarrow 0\%$) & 1M ($\downarrow 0\%$) \\ \hline
    \end{tabular}}
\end{subtable}

\vspace{0.1cm}

\begin{subtable}[t]{\textwidth}
    \centering
    \caption{Continuation: Choice of underrepresented and overrepresented groups for each dataset after filtering, used to measure the models' fairness.}
    \label{tab:datasets_2}
    \resizebox{\textwidth}{!}{
    \begin{tabular}{|c|c|c|}
    \hline
    \textbf{Dataset} & \textbf{Underrepresented Groups} & \textbf{Overrepresented Groups} \\ \hline
    Beauty& Austria, Egypt, Hungary, Portugal, Bulgaria & China, USA, France, South Korea \\ \hline
    % Grocery& Brazil, Greece, Belgium, Philippines, Indonesia, Costa Rica & USA, China, Italy, Canada, India \\ \hline
    Instruments & Nepal, Vietnam, Malaysia, Philippines, Denmark, Austria & China, USA, Taiwan, Canada \\ \hline
    % Video Games & Hong Kong, Canada & China, USA \\ \hline
    LastFM-2b & Netherlands, Japan, Brazil, Poland, Norway & USA, UK, Canada, France \\ \hline
    MovieLens-1M & Ireland, Japan, Korea, Hong Kong, China & USA, UK, France \\ \hline
    \end{tabular}}
\end{subtable}

\end{table}

\vspace{-0.7cm}
\subsection{Baselines}

To evaluate our model's performance, we select four baselines: Matrix Factorization (MF) \citep{koren2009matrix}, anchor-based collaborative filtering (ACF) \citep{barkan2021anchor}, ProtoMF \citep{melchiorre2022protomf}. Additionally, we include a benchmark for popularity bias mitigation, ZeroSum \citep{rhee2022countering}, by adding the score difference regularizer to the loss function of user and item side of ProtoMF. This allows for a fair comparison of its effectiveness with our approach.

Hyperparameter tuning is performed using the Ray library \citep{moritz2018ray} across 50 different seeds. For ProtoMF-based models, we first optimize the vanilla ProtoMF configuration. Fixing these hyperparameters, we then separately optimize our additional parameters to reduce the hyperparameter search space and ensure computational feasibility.

\subsection{Metrics}
\begin{itemize} \item \textbf{Performance Metrics}: We employ two standard accuracy metrics of Hit Ratio (HR@10) and Normalized Discounted Cumulative Gain (NDCG@10) to assess the performance of the models as \cite{he2017neural, melchiorre2022protomf} have previously done.

\item \textbf{Fairness Metrics}: \begin{itemize} \item \textit{Per-group Item Average Ranking}: Ranking plays a critical role in fairness, as user interactions vary based on item positions, as explored by \cite{ursu2018power}. We report the average ranking of items for underrepresented ($\mu(r_{under})$) and overrepresented groups ($\mu(r_{over})$) across model outputs of 100 items.
    \item \textit{Long-Tail Average Ranking}: We also report the average ranking of long-tail items for all users. Long-tail items are defined as the 10\% of items with the \textit{fewest} interactions, based on the logarithmic interaction count discussed by \citet{salganik2024fairness}.
\end{itemize}
\end{itemize}

\section{Results} % DONE
\label{sec:results}

\begin{figure}[ht]
    \centering
    \begin{subfigure}{0.49\textwidth}
        \centering
        \includegraphics[width=\textwidth]{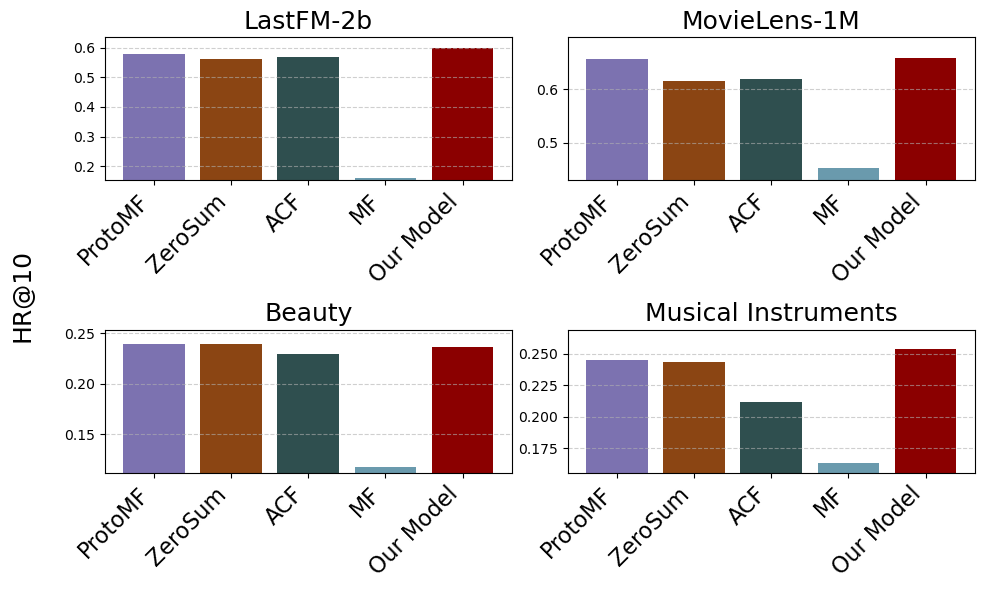}  % Adjust the width to your needs
        \caption{Hit Rate @ 10}
        \label{fig:hr10}
    \end{subfigure}
    \hfill
    \begin{subfigure}{0.49\textwidth}
        \centering
        \includegraphics[width=\textwidth]{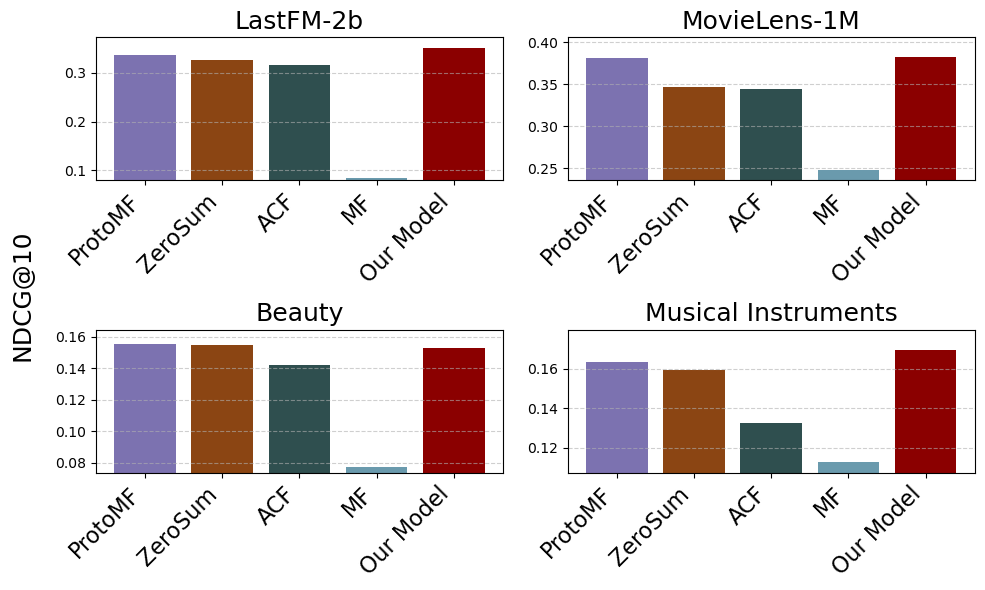}  % Adjust the width to your needs
        \caption{NDCG @ 10}
        \label{fig:ndcg10}
    \end{subfigure}

    \begin{subfigure}{0.49\textwidth}
        \centering
        \includegraphics[width=\textwidth]{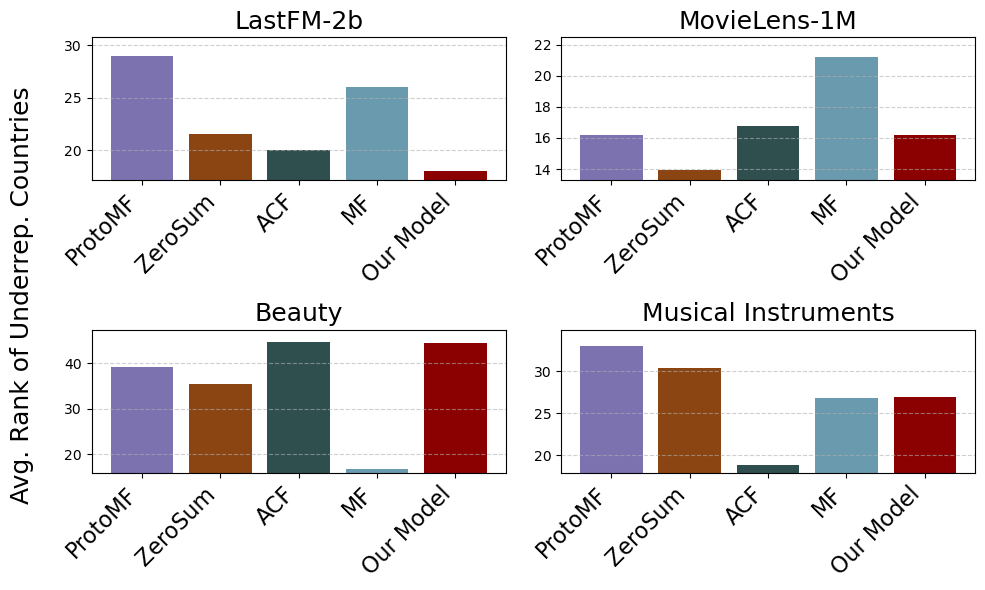}  % Adjust the width to your needs
        \caption{Avg. Rank for Long-tail Items}
        \label{fig:long_tail}
    \end{subfigure}
    \hfill
    \begin{subfigure}{0.49\textwidth}
        \centering
        \includegraphics[width=\textwidth]{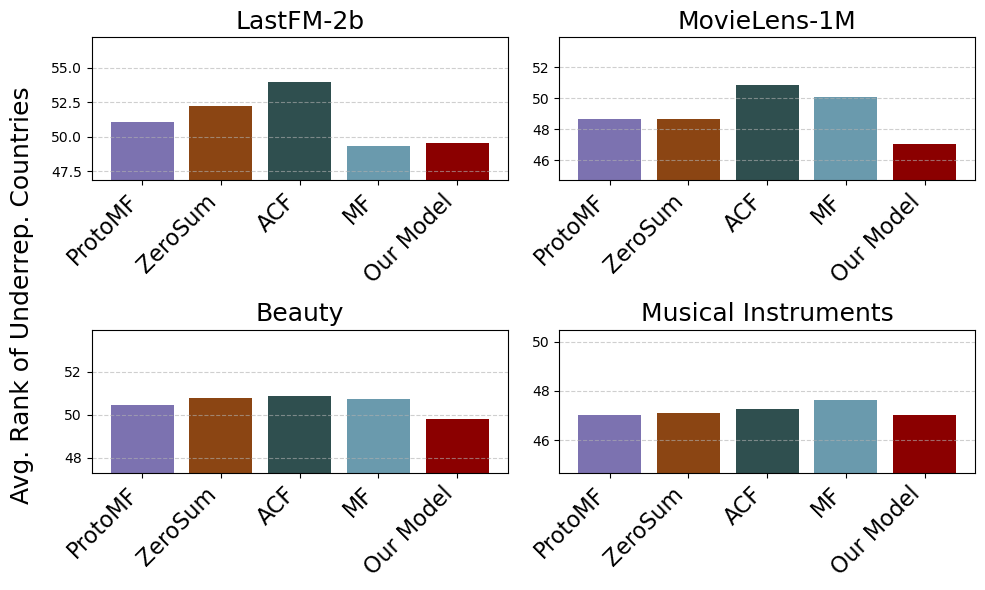}  % Adjust the width to your needs
        \caption{Avg. Rank for Underrep. Countries}
        \label{fig:underrepresented}
    \end{subfigure}
    
    \caption{The first group (a, b) presents the performance metrics. The second group (c, d) presents the fairness metrics. Our model with the highest HR@10 is chosen.}
    \label{fig:bar_plots}
\end{figure}

Our findings, illustrated in Figures \ref{fig:hr10} and \ref{fig:ndcg10}, demonstrate that our model surpasses MF and ACF in terms of performance metrics and achieves competitive results compared to ProtoMF and ZeroSum. Specifically, on the LastFM dataset, our model attains a HitRatio@10 of 0.600, marking a 3\% improvement over ProtoMF's score of 0.581.

Crucially, this performance enhancement is accompanied by significant improvements in fairness metrics. By applying our regularization techniques to promote a more uniform distribution of prototypes within the embedding space, we observe a 38\% reduction in the average ranking of long-tail items on the LastFM dataset (see Figure \ref{fig:long_tail}). Additionally, as indicated in Figure \ref{fig:underrepresented}, our model more effectively promotes items from underrepresented groups (refer to  Table \ref{tab:datasets_2} for group definitions). For instance, on the MovieLens dataset, there is a 3.4\% reduction in the average ranking for items from underrepresented countries.

\begin{figure}[h]
  \centering
  \hspace{-.65cm}
  \includegraphics[scale=0.2098]{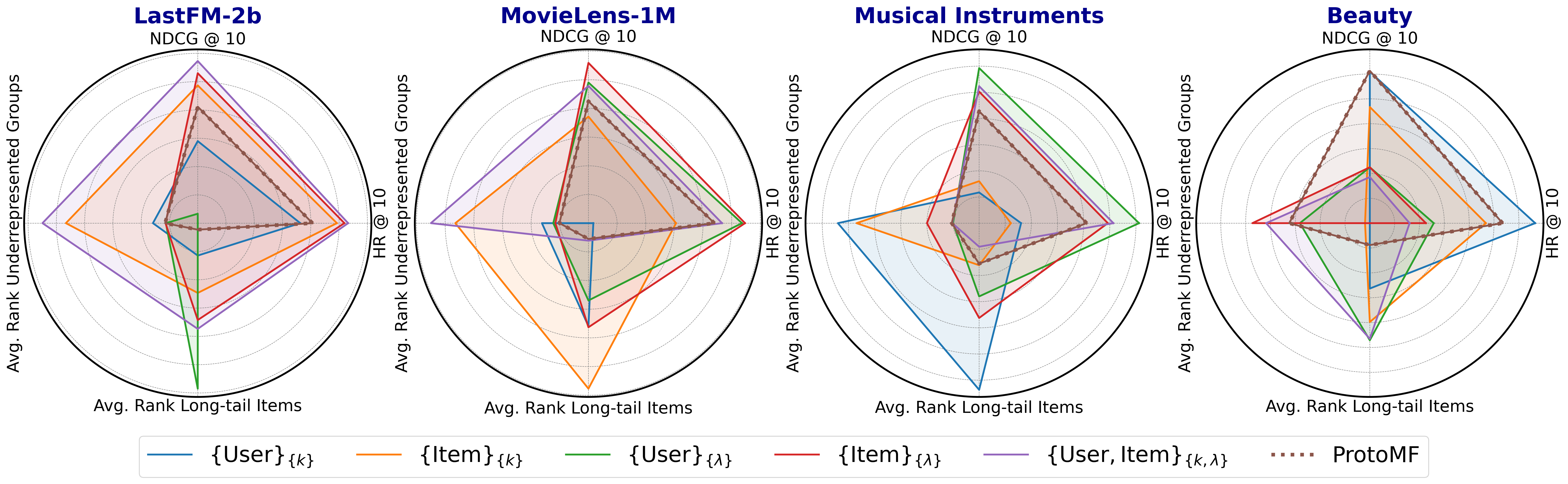} 
  % \caption{Spider plots showing the performance of four models ($\{\text{item}\}_\{k\}$, $\{\text{item\}}_\{\lambda\}$, $\{\text{user, item}\}_{\{k, \lambda\}}$, ProtoMF) across different datasets. Metrics include HR@10, NDCG@10, (negated) average underrepresentation, and (negated) long-tail ranking. Values are normalized per dataset, with higher values indicating better performance or fairness.}

  \caption{Results obtained for our various model configurations. All values are  normalized per dataset, with higher values indicating better performance or fairness.}

  \label{fig:spider_plots}
\end{figure}

Moreover, our method offers flexibility in balancing performance and fairness based on specific requirements. As depicted in Figure \ref{fig:spider_plots}, different configurations of our model result in varying levels of performance and fairness. For example, on the MovieLens dataset (see Table \ref{tab:all_experiment_results}), the configuration denoted as the \textcolor{violet}{$\{\text{User, Item}\}_{\{k, \lambda\}}$} model achieves an HR@10 of 0.657, slightly lower than the \textcolor{BrickRed}{$\{Item\}_{\{\lambda\}}$}  model's HR@10 of 0.660, but provides an 18\% lower average ranking for underrepresented groups, indicating enhanced fairness.

\begin{figure*}[htbp]
    \vspace{-0.5cm} % Reduce unnecessary vertical spacing
    \centering
    \begin{subfigure}[b]{0.44\linewidth} % Adjusted width for better spacing
        \centering
        \includegraphics[width=0.8\linewidth]{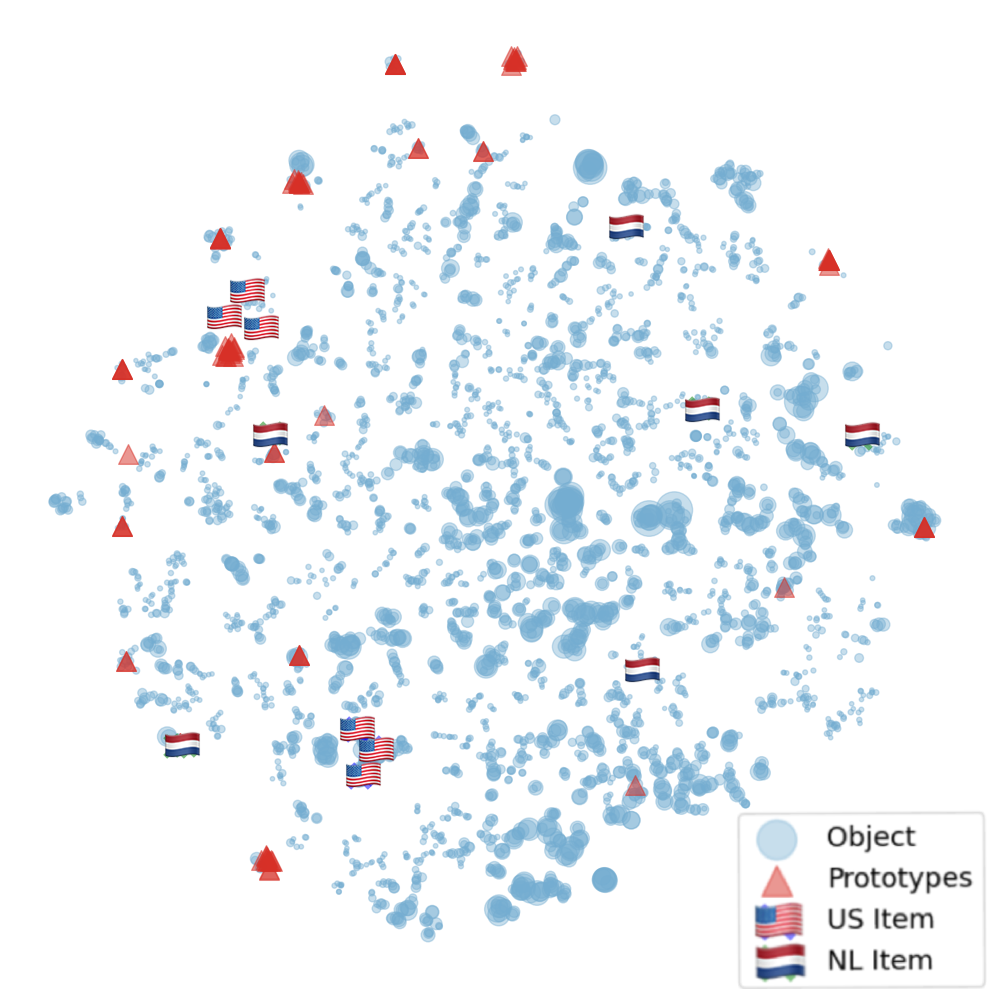}
        \caption{Prototypes cluster more around US items.}
        \label{fig:item_interactions_b}
    \end{subfigure}
    \hfill % Use flexible spacing instead of \hspace
    \begin{subfigure}[b]{0.44\linewidth} % Adjusted width
        \centering
        \includegraphics[width=0.8\linewidth]{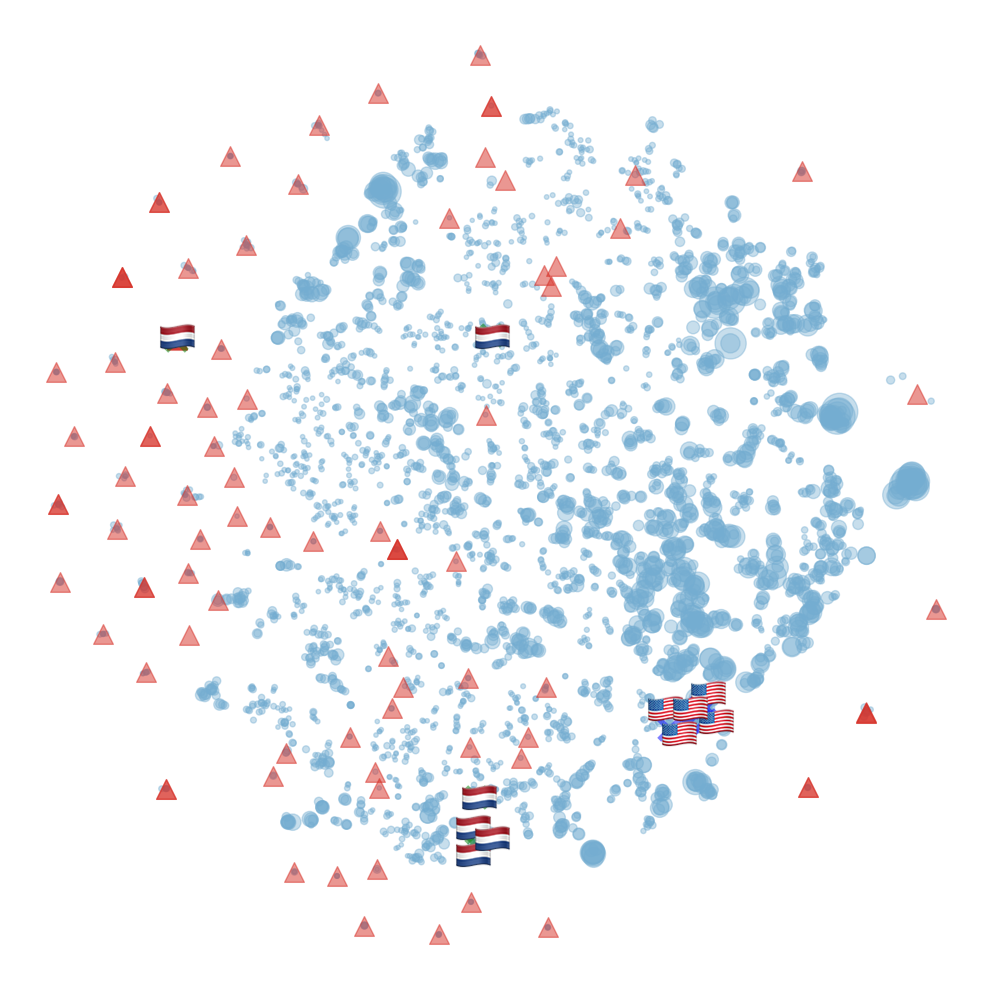}
        \caption{Prototypes are more evenly distributed. NL items are grouped.}
        \label{fig:item_interactions_c}
    \end{subfigure}
    \caption{t-SNE visualizations of the item embedding space for (a) ProtoMF and (b) $\{\text{User-Item}\}_{\{k, \lambda\}}$ models. NL and US items are respectively from under and over-represented countries.}
    \label{fig:embeddings}
\end{figure*}

In addition to enhancing fairness, our approach of distributing prototypes within the embedding space facilitates the allocation and grouping of more items from underrepresented groups. Figure \ref{fig:embeddings} illustrates the differences in the embedding space before and after applying the regularizer.

\section{Discussion}
\label{sec:discussion}

\noindent\textit{\textbf{Item Side: Synergy of k-Filtering and Prototype Distribution:}} Applying item k-filtering yielded significant improvements in ranking underrepresented items. By representing points with fewer but more relevant local neighbors, k-filtering enhances the representation of items, ensuring that prototypes better reflect the true data distribution. This technique improves the model's ability to accurately rank items from less represented countries without adversely affecting—and sometimes even improving—the ranking of overrepresented items.

Conversely, incorporating item-level regularization $\text{Item}_\lambda$ proved most effective for promoting long-tail items. The regularizer forces prototypes to spread out in the embedding space, encouraging the exploration of regions that would otherwise remain underutilized (see Figure \ref{fig:item_interactions_b}. However, this approach can lead to a decrease in HitRatio, as spreading prototypes may result in noisier representations embedded in sparsely populated regions. This trade-off highlights the limitations of solely applying prototype spreading.

Interestingly, the combination of item k-filtering and prototype regularization offered the best overall performance. This synergy balances refining representations of underrepresented items and encouraging diversity in the top recommended items. By integrating both methods, we mitigate the decrease in HitRatio caused by prototype spreading alone while still promoting diversity. The result is better rankings for items from underrepresented countries and a more varied recommendation list, enhancing both fairness and user satisfaction.

\noindent\textbf{\textit{User Side, Limited Effects on Performance:}} Contrary to our expectations, user-based techniques did not improve performance and often resulted in worse outcomes compared to the baseline. Methods like regularization or k-filtering applied to user representations had little effect on performance. Moreover, regularizing user embeddings may degrade fairness because it introduces noise into the user representations without reducing the allocational harms.

One possible explanation is that our evaluation focuses on the country of origin of the items (allocational fairness) rather than the origin of the users (representational fairness). Therefore, applying regularization in the user embedding space does not directly increase the visibility of items belonging to the long-tail or underrepresented groups. This highlights the importance of applying regularization in the appropriate embedding space to effectively address specific fairness concerns. User-based regularization might be more effective in scenarios where representational harms are  the primary concern.

\noindent\textbf{\textit{Item Side: Inclusiveness in Interpretability:}} As depicted in Figure \ref{fig:embeddings}, we can observe the impact of prototype distribution on the ability of the model provide improved explanations for items belonging to the minority groups. In our model, as shown in Figure \ref{fig:item_interactions_c}, the prototypes become more evenly distributed across the item embedding space. This redistribution ensures that items from underrepresented cultures are represented by closer prototypes. As a result, the model can offer more \textit{culturally-aware} explanations for its recommendations. Table \ref{tab:explainability_examples} shows nearest prototype examples of a Japanese animation and an Aztec musical instrument: The prototypes associated with our model take up not only the category of the product but also its cultural proximity, making the recommendations more intuitive, representative, and inclusive.

% \vspace{-0.6cm}
% Add the following packages to your document preamble:
% \usepackage{booktabs}    % For professional table lines
% \usepackage{multirow}    % For multi-row cells
% \usepackage{graphicx}    % For scaling
% \usepackage{colortbl, xcolor} % For colors

% Define the custom heatmap color
\definecolor{heat_color}{RGB}{255, 165, 0} % Orange

\begin{table}[htbp]
\centering
\caption{Closest items to the five nearest prototypes for each product, listed in decreasing order of similarity (values in parentheses). Bolded items indicate those from the same country as the original product. Products are sourced from the Amazon Movies and Instruments datasets.}
\label{tab:explainability_examples}
% \tiny % Reduce the font size further
\setlength{\tabcolsep}{1.5pt} % Reduce column separation
\scalebox{0.505}{ % Scale the table
\begin{tabular}{@{}p{2.0cm} p{1.5cm} *{5}{c}@{}}
\toprule
\textbf{Product} & \textbf{Model} & \textbf{Proto. 1} & \textbf{Proto. 2} & \textbf{Proto. 3} & \textbf{Proto. 4} & \textbf{Proto. 5} \\ 
\midrule

\multirow{2}{2.0cm}{\makecell[l]{\textit{Princess Mo-}\\\textit{nonoke (JP)}}} 
 & Baseline 
 & The Lion King (1.58) (US)
 & \textbf{Pokémon} (1.51) (JP)
 & Aladdin (1.50) (US)
 & Small Soldiers (1.48) (US)
 & Toy Story 2 (1.47) (US) \\ 
\cmidrule(l){2-7}
 & Ours 
 & \textbf{Pokémon (1.64)} (JP)
 & \textbf{Akira (1.59)} (JP)
 & The Lion King (1.55) (US)
 & \textbf{Perfect Blue (1.52)} (JP)
 & The Iron Giant (1.31) (US) \\ 
\hline

\multirow{2}{2.0cm}{\makecell[l]{\textit{Aztec Death}\\\textit{Whistle} (CO)}} 
 & Baseline 
 & Flute (1.56) (CN)
 & Ocarina (1.53) (CN)
 & Drumset (1.47) (US)
 & Saxophone (1.40) (CN)
 & Trumpet (1.32) (US) \\ 
\cmidrule(l){2-7}
 & Ours 
 & Ocarina (1.61) (CN)
 & Flute (1.58) (CN)
 & \textbf{Screaming Whistle} (1.53) (CO)
 & Trumpet (1.43) (US)
 & \textbf{Tambora} (1.42) (TW) \\ 
\bottomrule
\end{tabular}
}
\end{table}

\section{Conclusion \& Future Directions}

\label{sec:conclusion}

We introduced enhancements to Prototype-based Matrix Factorization to mitigate demographic biases in RecSys, focusing on cultural overrepresentation. By implementing Prototype K-filtering and a Prototype-Distributing Regularizer, our model improves the representation of underrepresented items without compromising recommendation quality. Experimental results show that our approach achieves higher performance than baseline models and significantly reduces the average ranking of long-tail and underrepresented items. This promotes fairness and inclusiveness in recommendations and explanations, contributing to a more equitable user experience.

However, our study has several limitations. First, the theoretical justification for the effectiveness of k-filtering and the regularization techniques is not thoroughly established; more rigorous analysis is needed to fully understand their impact on performance and fairness. Second, while we observed enhancements in explainability, we did not conduct a systematic analysis of this aspect. Additionally, the significant reduction of data due to filtering for country information may affect the generalizability of our findings. Finally, by simplifying culture to country of origin, we overlook its complex and multifaceted nature that extends beyond national boundaries; culture is influenced by factors such as ethnicity, language, and religion. This oversimplification may limit the effectiveness of our bias mitigation strategies and fail to capture the diversity within and across countries.

% \begin{itemize}
    % \item Today:
    % \item \textcolor{red}{Address Flo's comments on the writing}
    % \item \textcolor{red}{Add dataset full names to the dataset subsection}
    % \item \textcolor{red}{Remove Video Games' dataset and its remnants throughout the text}
    % \item \textcolor{red}{Explain the $\{Item\}_{\{k\}}$ notation in the methodology.}

    % \item Today:
    % \item \textcolor{red}{Add ZeroSum To The Table Appendix }
    % \item \textcolor{red}{Add some information about the definition of culture in the intro}
    % \item \textcolor{red}{Edit bar plot \ref{fig:bar_plots} and replace instruments with grocery, make it more presentable and readable}
    % \item \textcolor{red}{remove grocery from the main paper and add it only to the table 3 of appendix}
    % \item     \textcolor{red}{address comments}
    % \item \textcolor{red}{fix the heatmap on the last table}
    % \item \textcolor{red}{double check all the figure legends, remove redundant captions}

    % \item \textcolor{red}{fix examples of the last table, add distances}
    % \item \textcolor{red}{reiterate related work, the added examples}
    % \item \textcolor{red}{finish addressing the abstract comments}
    % \item \textcolor{red}{Do full iterations on text with Flo}
    
    % \item \textcolor{red}{Add more examples (except LastFM) to the \ref{tab:prototype_comparison_inclusive_explainability}}
    % \item \textcolor{red}{Address golnoosh's comments on related work}
% \end{itemize}

\newpage
\clearpage
\appendix
\label{appendix}
% \newpage % Optional: Use this if you want to ensure a new page starts with the section.
\section{Detailed Results}
\vspace{-.6cm}
\begin{table}[H]
\centering
\scriptsize
\vspace{-0.3cm}
\renewcommand{\arraystretch}{0.9} % Adjust the row height
\caption{Results of Models Across All Datasets. Bold numbers show the best value with respect to each dataset. $\mu_{\text{over}}$ to investigate how each model treats the overrepresented groups and in order to compare it against the $\mu_{\text{under}}$. We also include all the results on the `Amazon Grocery and Gourmet Food' dataset to provide more depth in our experiments.}
\label{tab:all_experiment_results}
\resizebox{\textwidth}{!}{%
\begin{tabular}{l*{3}{c}|c*{4}{c}}

\toprule
Model & Dataset & $\uparrow$HR@10  & $\uparrow$NDCG@10  & $\downarrow\mu_{\text{under}}$ & $\mu_{\text{over}}$ & $\downarrow\mu_{\text{LT}}$\\
\midrule

ZeroSum & LastFM-2b & 0.564 & 0.327 & 52.233 & 44.733 & 21.543 \\
    \citeauthor{rhee2022countering}   & MovieLens-1M & 0.615 & 0.347 & 48.660 & 49.381 & 13.964\\
        & Amazon Beauty & 0.238 & 0.154 & 50.771 & 49.634 & 35.372 \\
        & Amazon Instruments & 0.243 & 0.159 & 47.099 & 49.997 & 30.416 \\
        & Amazon Grocery & 0.261 & 0.170 & 50.582 & 49.088 & 18.413 \\

\midrule
ProtoMF & LastFM-2b & 0.581 & 0.337 & 51.106 & 44.300 & 29.008 \\
\citeauthor{melchiorre2022protomf}& MovieLens-1M & 0.656 & 0.381 & 48.674 & 49.582 & 16.215 \\
& Amazon Beauty & 0.239 & 0.155 & 50.425 & 49.646 & 39.129 \\
& Amazon Instruments & 0.245 & 0.164 & 47.012 & 50.002 & 32.965 \\
& Amazon Grocery & 0.259 & 0.168 & 50.350 & 49.132 & 27.991 \\

% & Video Games & 0.134 & 0.084 & 52.096 & 50.135 & 28.712 \\
\midrule
ACF & LastFM-2b & 0.569 & 0.317 & 53.970 & 43.458 & 20.009 \\
\citeauthor{barkan2021anchor}& MovieLens-1M & 0.618 & 0.345 & 50.891 & 53.923 & 16.783 \\
& Amazon Beauty & 0.229 & 0.142 & 50.885 & 49.615 & 44.690 \\
& Amazon Instruments & 0.211 & 0.132 & 47.250 & 49.924 & 18.849 \\
& Amazon Grocery & 0.250 & 0.159 & 50.530 & 49.137 & 24.343 \\

% & Video Games & 0.095 & 0.095 & 51.899 & 50.161 & 19.455 \\
\midrule
MF & LastFM-2b & 0.159 & 0.083 & 49.299 & 49.755 & 25.991 \\
\citeauthor{koren2009matrix}& MovieLens-1M & 0.452 & 0.248 & 50.098 & 49.340 & 21.201 \\
& Amazon Beauty & 0.117 & 0.077 & 50.711 & 49.595 & 16.660 \\
& Amazon Instruments & 0.163 & 0.113 & 47.620 & 49.926 & 26.825 \\
& Amazon Grocery & 0.132 & 0.086 & 50.338 & 49.208 & 21.541 \\

% & Video Games & 0.106 & 0.067 & 51.771 & 50.100 & 11.267 \\
\midrule
\midrule

$\{\text{user}\}_{\{k\}}$ & LastFM-2b & 0.574 & 0.326 & 50.952 & 44.587 & 26.136 \\
& MovieLens-1M & 0.640 & 0.365 & 48.452 & 50.956 & 13.330 \\
& Amazon Beauty & \textbf{0.240} & \textbf{0.155} & 51.846 & 49.573 & 28.715 \\
& Amazon Instruments & 0.234 & 0.152 &\textbf{ 45.949} & 50.077 & \textbf{9.865 }\\
& Amazon Grocery & \textbf{0.262} &\textbf{ 0.170} & 50.765 & 49.118 & 19.098 \\

% & Video Games & 0.127 & 0.075 & 52.440 & 50.043 & 20.045 \\
\midrule
$\{\text{item}\}_{\{k\}}$ & LastFM-2b & 0.594 & 0.344 & 49.849 & 43.852 & 21.995 \\
& MovieLens-1M & 0.651 & 0.379 & 47.357 & 49.898 & \textbf{11.263} \\
& Amazon Beauty & 0.238 & 0.154 & 51.762 & 49.588 & 20.693 \\
& Amazon Instruments & 0.233 & 0.154 & 46.123 & 50.044 & 32.674 \\
& Amazon Grocery & 0.259 & 0.167 & 50.560 & 49.116 & \textbf{16.457} \\

% & Video Games & 0.113 & 0.066 & 51.151 & 50.194 & 15.924 \\
\midrule
\{$\text{user}\}_{\{\lambda\}}$ & LastFM-2b & 0.520 & 0.302 & 51.115 & 45.007 & 21.391 \\
& MovieLens-1M & 0.660 & 0.383 & 48.596 & 51.154 & 14.179 \\
& Amazon Beauty & 0.237 & 0.153 & 50.613 & 49.630 & 16.327 \\
& Amazon Instruments & \textbf{0.254} & \textbf{0.170} & 47.024 & 50.012 & 26.963 \\
& Amazon Grocery & 0.262 & 0.169 & 50.564 & 48.855 & 28.072 \\

% & Video Games & 0.123 & 0.073 & 52.104 & 50.180 & 30.904 \\
\midrule
$\{\text{item}\}_{\{\lambda\}}$ & LastFM-2b & 0.598 & 0.348 & 51.111 & 44.085 & 18.995 \\
& MovieLens-1M & \textbf{0.660} & \textbf{0.386} & 48.622 & 51.089 & 13.293 \\
& Amazon Beauty & 0.237 & 0.153 & \textbf{49.776} & 49.641 & 44.378 \\
& Amazon Instruments & 0.249 & 0.166 & 46.781 & 49.957 & 23.017 \\
& Amazon Grocery & 0.261 & 0.169 & \textbf{50.480} & 48.999 & 23.561 \\

% & Video Games & 0.125 & 0.075 & 52.060 & 50.173 & 14.899 \\
\midrule
$\{\text{user-item}\}_{\{k,\lambda\}}$ & LastFM-2b & \textbf{0.600} & \textbf{0.352} & \textbf{49.552} & 44.296 & \textbf{18.013} \\
& MovieLens-1M & 0.657 & 0.383 & \textbf{47.048} & 49.571 & 16.158 \\
 & Amazon Beauty & 0.236 & 0.153 & 50.018 & 49.643 & \textbf{16.684} \\
& Amazon Instruments & 0.250 & 0.167 & 47.017 & 49.993 & 36.062 \\
& Amazon Grocery & 0.261 & 0.168 & 50.568 & 49.047 & 26.788 \\

% & Video Games & 0.126 & 0.075 & 52.269 & 50.163 & 21.299 \\

\bottomrule
\end{tabular}%
}
\end{table}

\bibliographystyle{ACM-Reference-Format}
\bibliography{main}

\end{document}